\documentclass[10pt,a4paper,onecolumn]{article}
\usepackage{graphicx}
\usepackage{color}
\begin{document}
\title{Realistic thermal heat engine model and its generalized efficiency}

\author{M. Ponmurugan \\
Department of Physics, School of Basic and Applied Sciences, \\
Central University of Tamilnadu, Thiruvarur - 610 005,  \\
Tamilnadu, India. e-mail:ponphy@cutn.ac.in}



\maketitle

\begin{abstract}
We identify  a  realistic model of thermal heat engines and obtain the generalized efficiency,
$\eta= 1- \left(\frac{T_c}{T_h}\right)^{1/\delta}$, where 
$\delta=1+\frac{1}{\gamma}$ and $\gamma$ is the ratio of thermal heat 
capacities of working substance at two thermal stages of the hot heat reservoir 
temperature, $T_h$ and the cold heat reservoir temperature, $T_c$.
We find that the observed efficiency of practical heat 
engines satisfy the above generalized efficiency with $1/\delta=0.35594$ $\pm$ $0.07$.  
The Curzon-Ahlborn efficiency, $\eta_{CA}=1-\left(\frac{T_c}{T_h}\right)^{1/2}$ is obtained for the 
symmetric case, $\gamma=1$. The generalized efficiency approaches the Carnot efficiency, $\eta_C=1-\frac{T_c}{T_h}$, in the asymmetric limit, $\gamma \to \infty$. 
\end{abstract}


{\it Introduction--} Thermodynamic studies of heat engines often focus on finding a realistic model
whose efficiency, $\eta=W/Q_h$, should match with the efficiency of practical 
heat engines \cite{Hoffmann,Espo,rgen1}. Here, $W$ is the work performed in a given cycle, in which $Q_h$ amount of heat absorbed from 
the hot reservoir at a higher temperature $T_h$ and $Q_c$ amount heat delivered to the cold 
reservoir at lower temperature $T_c$. Power delivered by the cyclic heat engine is $P=W/\tau$, 
where $\tau$ is the total time taken to complete the given cycle. The efficiency of realistic heat 
engines are bounded below the idealistic zero power Carnot engine efficiency, $\eta_C=1-T_c/T_h$.

The earliest works on the heat engine model entirely based on  the endoreversible approximation \cite{Hoffmann} attempts to obtain the efficiency at non-zero (maximum) power. This is the so called Curzon-Ahlborn efficiency, $\eta_{CA}=1-\left(\frac{T_c}{T_h}\right)^{1/2}$ \cite{novikov,curzon} with the exponent $1/2$ in the temperatures ratio, $T_r=T_c/T_h$ \cite{Hoffmann}, which is closely matches with the efficiency of realistic heat engines \cite{curzon,eff1by3}. There is a general belief that $\eta_{CA}$ should be the efficiency of the realistic heat engines \cite{geneff,avatar}.  However, it has been pointed out that the efficiency of heat engines can also be obtained
with other values of exponent in $T_r$ \cite{eff1by3,eff1by4}.
In particular, the exponent $1/3$ is obtained for the efficiency at maximum power of endoreversible model of the 
coupled heat engine \cite{eff1by3} and the exponent $1/4$ is obtained  for the irreversible 
Brownian heat engine model \cite{eff1by4}.

Recent studies on the efficiency of Carnot engine whose working substance is a cosmological model of variable generalized Chaplygin gas and polytropic gas, also showed that there should be an exponent in $T_r$ which is different from the above observed values. 
The exponent associated with the free parameters given in the different equations of state of the cosmological gas models are in general not equal to one even for reversible Carnot cycle efficiency \cite{malav,poly}.
This urge us to identify a suitable model for heat engines whose efficiency (irrespective of different
heat transfer processes and optimization conditions) provides a generalized 
value of the exponent in $T_r$, which closely matches with the observed efficiencies of practical thermal heat engines \cite{Espo,geneff,Johal}. 
 
Most of the heat engine studies discussed above are based on the assumption that the temperature of the working substance does not change during the heat transfer process though considering the heat exchange at the boundaries. It has been observed that the heat exchange at the boundary of the heat baths significantly affect the temperature of the working substance \cite{rgen1,lef1,lef2,guo}. Further, the performance of the heat engine also depends on the 
nature of heat capacities of the heat reservoirs and the working substance \cite{rgen1,lef1,lef2,guo,rgen2}. In this paper,
we  consider the fact that the thermal heat capacities of the working substance is in general different 
at high and low temperatures \cite{rgen1}. We utilize the difference in thermal heat capacities of working substance at two thermal stages and calculate the local equilibrium temperature of the working substance, which is then used to find out the efficiency of the realistic heat engine model.


{\it Model--} The important working principle of realistic heat devices is that the system does not
evolve far from the required operating condition \cite{hern}. In particular, the ability to get retained in the stationary state defines the operation regime of the cyclic process which maintains the control and stability of the same \cite{hern}. 
In our model, the working substance kept between the hot and cold equilibrium reservoirs is initially at an arbitrary temperature.  The working substance reaches a local equilibrium (stationary or steady state) at  
temperature $\theta$ by absorbing $\tilde{Q}_h$ amount of heat from the hot  heat reservoir 
and releasing  $\tilde{Q}_c$ amount of heat to the cold heat reservoir. Our model is based on the 
assumption that the arbitrary temperature of the working substance is initially at $T_h$ and finally
reaches $T_c$ during the heat exchange processes. Once the working substance attains a local equilibrium  
temperature $\theta$, it remains locally in equilibrium thereafter and undergoes only reversible processes. In the present model, the heat engine as a whole is considered to be in non-equilibrium state.  However, the working substance should be remained locally in equilibrium once it reaches the temperature $\theta$. The condition imposed to achieve this is that the total entropy production of the working substance during the entire stage should be zero. In this aspect, our model is different from the endoreversible model of various heat engines studied earlier \cite{Hoffmann}. In endoreversible thermodynamics, the system is considered as connected parts of internally reversible (endoreversible) subsystems and the energy exchange between them take place in an irreversible manner \cite{Hoffmann}.

\begin{figure}
\centering
\includegraphics[scale=0.4]{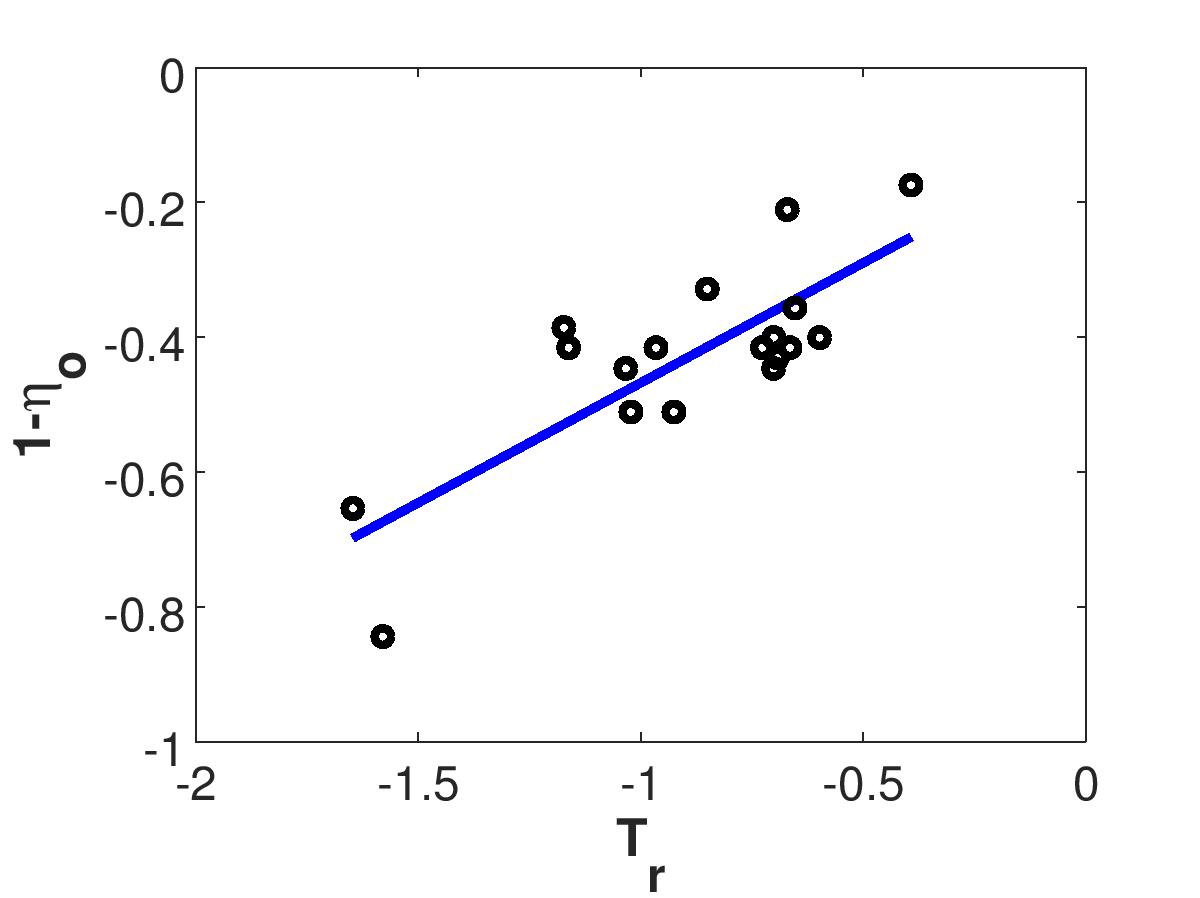}
\caption{
Log - Log plot of $1-\eta_o$ versus temperature ratio $T_r=T_c/T_h$. 
$\eta_o$ is the observed efficiency of thermal power plants. 
Linear fit (solid line) gives the slope 0.35594 $\pm$ $0.07$.
}\label{ebfig}
\end{figure}

To obtain the local equilibrium temperature of the working substance, we consider that the heat transfer between the working substance and the heat reservoirs take place in two steps. First, it absorbs $\tilde{Q}_h$ amount of heat from the hot bath in a finite time $\tau_1$. 
The arbitrary temperature $T(t)$ of the working substance is assumed  initially at the temperature of the 
hot reservoir $T_h$ and it reaches the local equilibrium temperature $\theta$ $(< T_h)$ during the time $\tau_1$. 
Second, the working substance releases remaining heat $\tilde{Q}_c$  
to the cold heat reservoir in a finite time  $\tau_2$  by  maintaining 
its local equilibrium  temperature $\theta$ $(> T_c)$. The arbitrary temperature of the working substance 
is now assumed to reach the cold reservoir temperature $T_c$ during the time interval $\tau_2$. 
In order to ensure that the working substance maintains the local equilibrium 
temperature $\theta$, we impose the condition that the total entropy production of the working substance 
should be zero during the total time interval $\tau=\tau_1+\tau_2$. Such as,
\begin{eqnarray}\label{entprd}
 \int_0^{\tau_1} \frac{d\tilde{Q}_h}{T(t)} - \int_{\tau_1}^{\tau_2} \frac{d\tilde{Q}_c}{T(t)} = 0.
\end{eqnarray}
The convention we used here is that the heat flow in to the system is positive.
In order to calculate $\theta$ in terms of $T_h$ and $T_c$, we utilize the concept of realistic system that 
the thermal heat capacities of the  working substance, $C_s$, is different at two 
thermal stages of reservoirs \cite{rgen1}. We rewrite Eq.(\ref{entprd}) in terms of heat capacity 
$C_s^i=d\tilde{Q}_i/dT$, $(i:h,c)$ of working substance when it exchange  heat 
between the hot reservoir at temperature $T_h$ and cold reservoir at temperature $T_c$ as 
\begin{eqnarray}\label{entprd1}
 \int_0^{\tau_1} \frac{C_s^h dT}{T} - \int_{\tau_1}^{\tau_2} \frac{C_s^c dT}{T} = 0  \\ \nonumber
  C_s^h \int_{T_h}^{\theta} \frac{dT}{T} - C_s^c \int_{\theta}^{T_c} \frac{dT}{T} = 0.
\end{eqnarray}
Solving the above equation, one can get
\begin{eqnarray}\label{eqtemp}
\theta=T_c^{\frac{\gamma}{\gamma+1}}T_h^{\frac{1}{\gamma+1}},
\end{eqnarray}
where $\gamma=C_s^c/C_s^h$. It should be noted that similar kinds of the above relation has 
been observed in different contexts \cite{lef1,lef2,guo,rgen2} of the Carnot and non-Carnot heat engines.
In contrast, working condition of our model is 
based on the assumption that the local equilibrium temperature $\theta$ (Eq. \ref{eqtemp}) 
of working substance obtained in a time 
interval $\tau$ is much lesser than the infinitesimally small time delay in 
which the realistic heat engine can start any cyclic operations. Under this assumption, 
the heat engine can operate in finite time arbitrary cycle in a (locally equilibrium) reversible manner by performing  
useful work $W=Q_h-Q_c$  while absorbing $Q_h$ amount of heat from the hot reservoir 
at tempeaturte $T_h$ and  releasing $Q_c$ amount of heat to the surrounding 
with the effective working substance temperature $\theta$. Under this arbitrary  
cyclic process the total change in entropy of the system is
\begin{eqnarray}\label{dels}
\Delta S=\frac{Q_h}{T_h}-\frac{Q_c}{\theta}=0  \\ \nonumber
 \frac{Q_c}{Q_h} = \frac{\theta}{T_h}.
\end{eqnarray}
Even though the working substance of a heat engine operating between two temperatures 
$T_h$ and $T_c$, because of the arbitrary heat transfer between the system and the surroundings, 
we stress here that the realistic heat engine as a whole working 
only between the hot reservoir temperature $T_h$ and (local) equilibrium working substance 
temperature $\theta$.
By using Eq.(\ref{eqtemp}) and Eq.(\ref{dels}), the efficiency of the heat engine becomes,
\begin{eqnarray}\label{eff}
 \eta&=&\frac{W}{Q_h}=1-\frac{Q_c}{Q_h}=1-\frac{\theta}{T_h} \\ \nonumber
   &=&1-\left(\frac{T_c}{T_h}\right)^{\frac{1}{\delta}},
\end{eqnarray}
where $\delta=1+\frac{1}{\gamma}$. Thus, we obtained the generalized efficiency of the
realistic heat engine. In particular, the Curzon-Ahlborn efficiency, 
$\eta_{CA}=1-\left(\frac{T_c}{T_h}\right)^{1/2}$ is obtained in the symmetric case of $\gamma=1$ and  
the efficiency approaches the Carnot efficiency, $\eta_C=1-\frac{T_c}{T_h}$ in the asymmetric 
limit of  $\gamma \to \infty$.  Eq. (\ref{eff}) can written in terms of 
$\eta_C$ as $\eta=1-\left(1-\eta_C\right)^{\frac{1}{\delta}}$. 
One can also see that the efficiency at maximum power of the weak dissipation 
non-Carnot heat engines \cite{guo} under symmetric dissipation condition has been observed 
in the above relation for $\gamma=1/2$.
With small temperature difference,
$\eta$ can be expanded in terms of $\eta_C$ and obtained the universal form of the relation 
\begin{eqnarray}\label{uni}
 \eta&=&\frac{1}{\delta} \eta_C + \frac{\delta-1}{2\delta^2} \eta_C^2 +\frac{(\delta-1)(2\delta-1)}{6\delta^3} \eta_C^3+O(\eta_C^4).
 \end{eqnarray}

We emphasis here that the generalized 
efficiency obtained in our model is generally valid for realistic heat engines operating 
under arbitrary operative conditions.
In order to strengthen our arguments,  we have plotted the observed efficiencies of the 
thermal plants \cite{Espo,geneff,Johal} 
versus temperature ratio $T_r$. Figure.1 shows a linear trend and the slope gives the 
value of $1/\delta = 0.35594$ $\pm$ $0.07$.  This shows  that the observed efficiencies of practical heat 
engines in general satisfy the above generalized efficiency with $1/\delta=0.35594$ $\pm$ $0.07$ 
which is much less than the Curzon-Ahlborn efficiency temperature ratio exponent of $0.5$.

{\it Conclusion --}
	We formulated a thermal heat engine model and  obtained the efficiency of  
realistic heat engines with a generalized exponent $(1/\delta)$ in the temperature ratio between 
the cold and hot reservoirs temperature. We found that the value of the exponent  
$1/\delta=0.35594$ $\pm$ $0.07$ in the temperature ratio of real power plants which 
is in general different from the usually expected Curzon-Ahlborn efficiency 
exponent of $0.5$.  However, the Curzon-Ahlborn efficiency obtained in the 
symmetric case of $\gamma=1$. We also obtained the efficiency at maximum power 
of the endoreversible Carnot (coupled) heat engines \cite{Hoffmann,eff1by3}
and symmetric (weak) dissipation 
non-Carnot heat engines \cite{guo} in the asymmetric case of $\gamma=1/2$.
The generalized efficiency approaches the $\eta_C$ in the asymmetric limit, $\gamma \to \infty$. 
We also expanded $\eta$ in terms of $\eta_C$ and 
obtained the generalized universal form of the efficiency.


\end{document}